\documentclass[prb,aps,twocolumn,showpacs,amssymb,superscriptaddress]{revtex4-1}
\usepackage{graphicx}
\usepackage{multirow,booktabs}
\usepackage{mathtools}
\usepackage{braket}
\usepackage{dcolumn}
\usepackage{bm}
\usepackage{color}
\usepackage{amsmath}
\usepackage{array}
\usepackage{float}
\usepackage[caption = false]{subfig}
\usepackage{verbatim}
\usepackage[colorlinks,hyperindex]{hyperref}
\hypersetup
	{	colorlinks,%
		citecolor=blue,%
		linkcolor=blue,%
		urlcolor=blue,%
	}

\begin{document}

\title{Chiral topological superconducting state with Chern number $\mathcal{C} =-2$ in Pb$_3$Bi/Ge(111)}

\author{Shuwen Sun}
\affiliation{International Center for Quantum Design of Functional Materials (ICQD), Hefei National Laboratory for Physical Sciences at Microscale (HFNL), and CAS Center for Excellence in Quantum Information and Quantum Physics, University of Science and Technology of China, Hefei, Anhui 230026, China}
\affiliation{Department of Physics, University of Science and Technology of China, Hefei, Anhui 230026, China}

\author{Wei Qin}
\thanks{Corresponding authors: \\ weiqin@utexas.edu; zhangzy@ustc.edu.cn}
\affiliation{International Center for Quantum Design of Functional Materials (ICQD), Hefei National Laboratory for Physical Sciences at Microscale (HFNL), and CAS Center for Excellence in Quantum Information and Quantum Physics, University of Science and Technology of China, Hefei, Anhui 230026, China}
\affiliation{Department of Physics, The University of Texas at Austin, Austin, Texas 78712,USA}

\author{Leiqiang Li}
\affiliation{International Center for Quantum Design of Functional Materials (ICQD), Hefei National Laboratory for Physical Sciences at Microscale (HFNL), and CAS Center for Excellence in Quantum Information and Quantum Physics, University of Science and Technology of China, Hefei, Anhui 230026, China}

\author{Zhenyu Zhang}
\thanks{Corresponding authors: \\ weiqin@utexas.edu; zhangzy@ustc.edu.cn}
\affiliation{International Center for Quantum Design of Functional Materials (ICQD), Hefei National Laboratory for Physical Sciences at Microscale (HFNL), and CAS Center for Excellence in Quantum Information and Quantum Physics, University of Science and Technology of China, Hefei, Anhui 230026, China}

\date{\today}

\begin{abstract}
Materials realization of chiral topological superconductivity is a crucial condition for observing and manipulating Majorana fermions in condensed matter physics. Here we develop a tight-binding description of Pb$_3$Bi/Ge(111), identified recently as an appealing candidate system for realizing chiral {\it p}-wave topological superconductivity [Nat. Phys. 15, 796 (2019)]. We first show that our phenomenological model can capture the two main features of the electronic band structures obtained from first-principles calculations, namely, the giant Rashba splitting and type-II van Hove singularity. Next, when the $s$-wave superconducting property of the parent Pb system is explicitly considered, we find the alloyed system can be tuned into a chiral topological superconductor with Chern number $\mathcal{C} = -2$, resulting from the synergistic effect of a sufficiently strong Zeeman field and the inherently large Rashba spin-orbit coupling. The nontrivial topology with $\mathcal{C} = -2$ is further shown to be detectable as two chiral Majorana edge modes propagating along the same direction of the system with proper boundaries. We finally discuss the physically realistic conditions to establish the predicted topological superconductivity and observe the corresponding Majorana edge modes, including the influence of the superconducting gap, Land\'{e} {\it g}-factor, and critical magnetic field. The present study provides useful guides in searching for effective {\it p}-wave superconductivity and Majorana fermions in two-dimensional or related interfacial systems. 
\end{abstract}

\maketitle

\section{introduction}
\label{sec:into}
	Majorana fermions that obey non-Abelian statistics may serve as the elemental entities for achieving topologically protected quantum computation\cite{A.Yu.Kitaev2003,Nayak2008}. In condensed matter systems, such exotic quasiparticles were first proposed to exist in the $\nu=5/2$ fractional quantum Hall state\cite{Moore1991}, subsequently also in chiral {\it p}-wave topological superconductors (TSCs)\cite{Read2000,Ivanov2001}. These conceptual developments have motivated extensive research efforts in their potential realizations, especially surrounding TSCs. Yet to date, it has been impossible to definitely detect an intrinsic {\it p}-wave superconducting state in realistic materials. One specific and widely explored system is the layered Sr$_2$RuO$_4$ \cite{Ishida1998}, exhibiting some signatures of {\it p}-wave pairing, but it has been actively debated since the initial report \cite{Mackenzie2003,Hicks2014,Steppke2017,Pustogow2019}.

In addition to searching for intrinsic chiral {\it p}-wave TSCs, alternative schemes have been put forward for realizing their effective counterparts via superconducting proximity effects \cite{Fu2008,Qi2010, Wang2015,Gorkov2001,Sau2010,Lutchyn2010,Alicea2010}. For example, the interface between the surface state of a three-dimensional (3D) topological insulator and a conventional {\it s}-wave superconductor was demonstrated to behave as a 2D chiral {\it p}-wave TSC \cite{Fu2008,Qi2010,Wang2015}. Similarly, a semiconductor with strong enough Rashba spin-orbit coupling (SOC) in proximity to a {\it s}-wave superconductor can also possess chiral {\it p}-wave superconductivity\cite{Gorkov2001,Sau2010,Lutchyn2010,Alicea2010}. In all of those schemes, a Zeeman splitting associated with an external magnetic field or intrinsic ferromagnetism is always required to lift the Kramers degeneracy at the time-reversal invariant $\Gamma$ point in the momentum space. Stimulated by these seminal proposals, signatures of topological superconductivity and Majorana fermions have been reported in several experimental setups, such as the semiconducting or metallic wires deposited on {\it s}-wave superconductors \cite{Mourik2012,Nadj-Perge2014,Albrecht2016,Feldman2017} and the topological or quantum anomalous Hall insulator-superconductor heterostructures\cite{Xu2015,Sun2016,He2017}. Although the observations of the quantized conductance \cite{Law2009,Chen2019,Zhu2020} and spin selective Andreev reflection \cite{He2014,Sun2016} have provided powerful evidence of Majorana zero modes, their most compelling and definitive smoking gun, namely the non-Abelian statistics, remain to be acquired, leaving plenty of room for further investigations\cite{Kayyalha2020}. 

Beyond the schemes described above that rely on proximity effects, there have also been efforts to explore the potential existence of Majorana zero modes in intrinsic TSCs, such as the transition metal doped topological insulators\cite{Fu2010,Sasaki2011,Hsieh2012} and doped iron selenide \cite{Yin2015,Zhang2018,Liu2018,Wang2018,Chen2019,Zhu2020}. Furthermore, it is highly desirable to achieve intrinsic topological superconductivity in 2D, as such systems are inherently superior to other architectures for braiding of Majorana fermions. To this end, various material platforms have been investigated as candidate 2D TSCs, for instance, the monolayered WTe$_2$ \cite{Wu2018,Fatemi2018,Sajadi2018}. In a more recent study, we have also showed that the combined effects of geometric phase and type-\uppercase\expandafter{\romannumeral2} Van Hove singularity (VHS) can give rise to intrinsic chiral {\it p}-wave superconductivity in the Pb$_3$Bi/Ge(111) system \cite{Qin2019}. Moreover, this system can harbor quantum spin Hall states under specific surface configurations, and the different quantum states can be converted into each other\cite{Li2020}. However, the intriguing topological properties of the superconducting quasiparticles and potential existence of the Majorana modes in Pb$_3$Bi/Ge(111) remain to be explored, especially under physically realistic conditions.

In this article, we develop a tight-binding model for Pb$_3$Bi/Ge(111) and perform quasiparticle band structure calculations to study its topological superconducting properties. First, we show that this phenomenological model can well reproduce the electronic bands around the Fermi level obtained from density functional theory (DFT) calculations, including the giant Rashba SOC and type-II VHS, which are two essential ingredients for realizing chiral {\it p}-wave superconductivity \cite{Qin2019}. Besides the time-reversal invariant $\Gamma$ and M points in the Brillouin zone (BZ), we also find a gapless Dirac cone protected by $C_{3v}$ lattice symmetry at the K point. When the {\it s}-wave superconductivity of the parent Pb system is considered, we reveal that the alloyed system can be tuned into a topological superconducting state with Chern number $\mathcal{C} = -2$ in the presence of a sufficiently strong Zeeman field. The nontrivial 2D bulk topology is further confirmed by edge state calculations, manifested as two chiral Majorana edge modes propagating along the same direction of the system with proper boundaries. Finally, we discuss the experimental realization of the predicted chiral topological superconductivity and Majorana edge modes under physically realistic conditions. With the practical superconducting gap, Land\'{e} {\it g}-factor, and critical magnetic field, Majorana edge modes can be achieved, whose penetration length into the bulk is on the order of tens of nanometers. These findings provide useful guides for probing chiral topological superconductivity in 2D or related interfacial systems.

This article is organized as followings. In Sec.~\ref{sec:model}, we present a phenomenological tight-binding model for the Pb$_3$Bi/Ge(111) system in detail. The central findings of this work are elaborated in Sec.~\ref{sec:res}, including the topological superconducting state with $\mathcal{C} = -2$ and the chiral Majorana edge modes. In Sec.~\ref{sec:discussion}, we discuss the potential experimental realization of these theoretical predictions. Finally, we summarize our results in Sec.~\ref{sec:dis}. 
 
\begin{figure}[tb]
	\includegraphics[width=1\columnwidth]{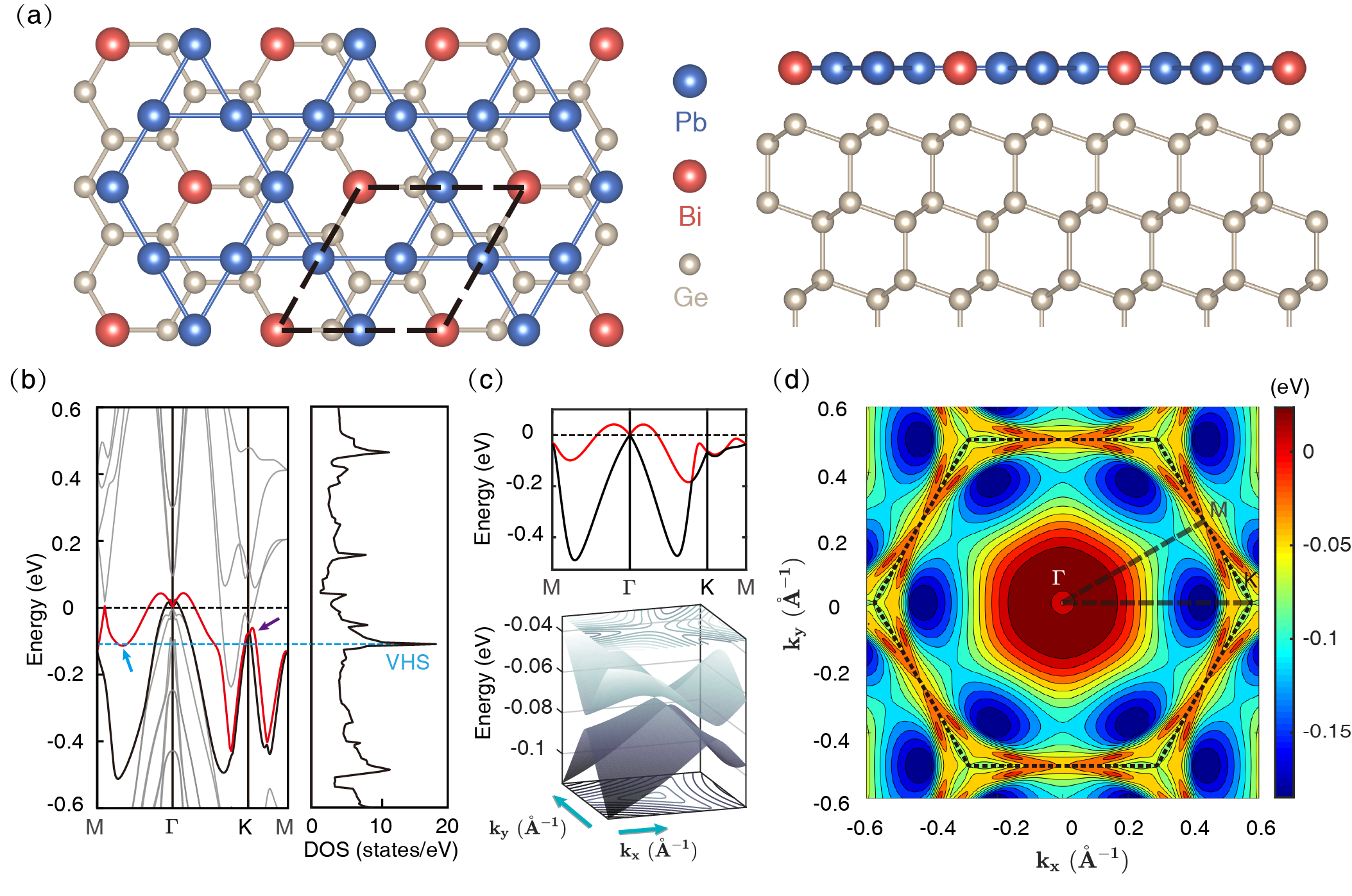}
	\caption{(a) Top (left) and side (right) views of the lattice structure of Pb$_3$Bi/Ge(111), where the dashed rhombus denotes the unit cell. (b) Band structure (left) and DOS (right) of Pb$_3$Bi/Ge(111) obtained from DFT calculations. In the left panel, the bold red and black curves around the Fermi level highlight the upper and lower Rashba bands, respectively, where the saddle point along the $\Gamma$-M direction is marked by a blue arrow, and the gapless Dirac cone at the K point is marked by a purple arrow. The gray bands are mainly contributed by the Ge(111) substrate. (c) Band structure obtained from the tight-binding model, in which we only plot the Rashba bands around the Fermi level, and the parameters used in this calculation are listed in Table~\ref{tb:Para}. Enlarged energy spectrum around the K point is depicted in the lower panel, and a gapless Dirac cone can be found. (d) Energy contour map of the upper Rashba band, where the dotted hexagon denotes the first BZ.}
	\label{fig1:structure}
\end{figure}

\section{Tight-binding model}
\label{sec:model}

As shown in Fig.~\ref{fig1:structure}(a), the lattice structure of a Pb$_3$Bi overlayer can be decomposed into a kagome sublattice of Pb atoms and a triangular sublattice of Bi atoms. When including the Ge(111) substrate, the heterostructure possesses $C_{3v}$ lattice symmetry with the central inversion symmetry broken by the substrate-induced potential. At the right hand side of Fig.~\ref{fig1:structure}(a), the side view of Pb$_3$Bi/Ge(111) is given, where we can see that the Pb and Bi atoms form an atomically flat monolayer.

Based on the recent DFT calculations \cite{Li2020}, the electronic band structures of this system is shown in Fig.~\ref{fig1:structure}(b). The large Rashba splitting of the bands around the Fermi level stems from the strong atomic SOC of the Pb and Bi atoms and effect of the substrate-induced inversion symmetry breaking. Another prominent feature is the type-\uppercase\expandafter{\romannumeral2} VHS contributed by the saddle-like band structure along the $\Gamma$-M direction in the upper Rashba band, leading to the logarithmically divergent density of states (DOS). These two features are essential for the emergence of chiral {\it p}-wave superconductivity emphasized in our earlier work \cite{Qin2019}. Here we first construct an effective tight-binding model aiming to reproduce the Rashba bands around the Fermi level, as highlighted in Fig.~\ref{fig1:structure}(b). The recent DFT calculations showed that the Rashba bands in Fig.~\ref{fig1:structure}(b) are dominated by contributions from the Pb and Bi atoms \cite{Li2020}. Therefore, we consider a simplified tight-binding model built on the Pb$_3$Bi overlayer as
\begin{equation} \label{Ht}
\begin{aligned}
H_0= &\sum_{i j}\left(\mu_1\delta_{i,j} - t_1\delta_{\langle i,j \rangle} -t_1' \delta_{\langle \langle  i,j \rangle \rangle}\right) a_{i}^\dagger a_{j} \\ 
-&\sum_{i l} \left(  t_2\delta_{\langle i,l \rangle} +t_2' \delta_{\langle \langle  i,l \rangle \rangle}\right) a_{i}^\dagger b_{l} + \sum_{l} \mu_2 b_{l}^{\dagger}b_{l} 
\end{aligned},
\end{equation}
where $a_i = (a_{i\uparrow},a_{i\downarrow})^{\text{T}}$ and $b_l = (b_{l\uparrow},b_{l\downarrow})^{\text{T}}$ are the annihilation operators on Pb and Bi atomic sites $i$ and $l$, respectively. Here $\mu_{1,2}$ are the on-site potentials for Pb and Bi atoms, $t_{1,2}$ and $t_{1,2}'$ are the nearest neighbor (NN) and next nearest neighbor (NNN) hopping amplitudes with 1(2) representing the Pb-Pb(Pb-Bi) hopping process, $\langle i,j \rangle$ and $\langle\langle i,j \rangle\rangle$ respectively denote that i and j are NN and NNN sites. We note that the above model only considers a single effective orbital with {\it s}-wave symmetry on each atomic site. Intriguingly, this maximally simplified model is able to capture the main features of the Rashba bands obtained from the DFT calculations. It should also be emphasized that, in the present study, we focus on explorations of the topological superconducting properties of the system mainly stemming from the Rashba SOC effect under an external magnetic field, but without explicitly invoking the type-\uppercase\expandafter{\romannumeral2} VHS.

Within this tight-binding model, the effect of substrate-induced inversion symmetry breaking can be described by a Rashba-type SOC Hamiltonian as
\begin{equation} \label{HR}
\begin{split}
H_{1}=-i\lambda_1&\sum_{\langle ij\rangle}a_{i}^\dagger(\bm{\sigma} \times \bm{d}_{ij})_za_{j} -i\lambda_{2} \sum_{\langle il\rangle}a_{i}^\dagger(\bm{\sigma} \times \bm{d}_{il})_z b_{l},
\end{split}
\end{equation}
where $\lambda_{1,2}$ are the Rashba coefficients for Pb-Pb and Pb-Bi hopping processes, 
$\bm{\sigma}$ is the Pauli vector, and $\bm{d}_{ij}$ is the in-plane vector pointing from site i to j. By performing Fourier transformation on $H= H_0+H_1$, we obtain a single-particle Hamiltonian in the momentum space written as $\mathcal{H}(\bm{k})$, and the electronic band structures through further diagonalization of $\mathcal{H}(\bm{k})$.

A typical result of the band structures is shown in Fig.~\ref{fig1:structure}(c), reproducing well the Rashba bands around the Fermi level obtained from the DFT calculations. The fitting parameters are listed in Table~\ref{tb:Para}, and the validity of these parameters can be justified qualitatively in the following three aspects. First, the hopping amplitude between two Pb atoms is larger than that between Pb and Bi atoms, consistent with the better metallic character of Pb. Secondly, the Rashba coefficient of Bi is larger than that of Pb. Thirdly, for electrons in the same atomic level of Pb and Bi, the onsite potential is higher for the atom with a smaller atomic number, consistent with $\mu_1>\mu_2$. In Fig.~\ref{fig1:structure}(c), the saddle point is evident in the upper Rashba band along the $\Gamma$-M direction. Due to the $C_{3v}$ symmetry, there are six such points within the first BZ as illustrated in Fig.~\ref{fig1:structure}(d), giving rise to a type-\uppercase\expandafter{\romannumeral2} VHS in the DOS \cite{Yao2015}. Therefore, we acquire both the strong Rashba splitting and type-\uppercase\expandafter{\romannumeral2} VHS characters, which further support the validity of this tight-binding model for the specific monolayered Pb$_3$Bi.

 \begin{table}[tb]
  	\caption{Physical parameters for the tight-binding model that gives the band structure shown in Fig.~\ref{fig1:structure}(c). All the parameters are in units of eV.}
  	\centering
  	\renewcommand\arraystretch{1.5}
  	\begin{tabular}{p{0.1\linewidth}<{\centering}p{0.1\linewidth}<{\centering} | p{0.07\linewidth}<{\centering} | p{0.1\linewidth}<{\centering} | p{0.1\linewidth}<{\centering}p{0.1\linewidth}<{\centering} | p{0.07\linewidth}<{\centering} | p{0.1\linewidth}<{\centering}}
  		\hline
  		\hline
  		\multicolumn{2}{c|}{\multirow{2}{*}{NN hopping}} & $t_1$ & 0.340	&\multicolumn{2}{c|}{\multirow{2}{*}{NNN hopping}} & $t_1'$ & 0.208	    \\
  		\cline{3-4}\cline{7-8}
  		\multicolumn{2}{c|}{} & $t_2$ & 0.280 & \multicolumn{2}{c|}{} & $t_2'$ & 0.120  \\ 
  		\cline{1-8}
  		\multicolumn{2}{c|}{\multirow{2}{*}{Rashba coefficient}} & $\lambda_1$	& 0.060	&\multicolumn{2}{c|}{\multirow{2}{*}{On-site potential}} & $\mu_1$	& 0.088	    \\
  		\cline{3-4}\cline{7-8}
  		\multicolumn{2}{c|}{} &  $\lambda_2$	& 0.160 & \multicolumn{2}{c|}{} & $\mu_2$	& -0.912  \\ 
  		\hline
  		\hline
  	\end{tabular}
  	\label{tb:Para}
\end{table}   

For the Rashba bands depicted in Figs.~\ref{fig1:structure}(b) and (c), gapless Dirac cones can be found at three high-symmetry points in the BZ. At $\Gamma$ and M, the Dirac points are protected by time-reversal symmetry, accompanied by conventional Rashba splittings away from those high-symmetry points. For the present system, K is not a time-reversal invariant point, but its gapless Dirac cone feature is protected by the C$_{3v}$ lattice symmetry \cite{Oguchi2009}. When enlarging the Rashba bands around the K point, a low-energy Dirac cone structure is clearly revealed, as depicted in the lower panel of Fig.~\ref{fig1:structure}(c). Signatures of the Dirac cone and Rashba-type vortical spin texture were also found around the K point in our recent DFT calculations\cite{Li2020}. In the remaining part of this work, we focus on the physics around the Dirac cone protected by the C$_{3v}$ lattice symmetry (located above the VHS), and explore the possibility of realizing topological superconducting states.

\section{Topological superconductivity}
\label{sec:res}
\subsection{Topological phase transition}
\label{sec:tpt}

According to earlier experimental observations, ultrathin Pb films and Pb$_{1-x}$Bi$_x$ alloys grown on Si(111) or Ge(111) substrates exhibit superconducting properties \cite{Ozer2007,Qin2009,Zhang2010,Sekihara2013,Yamada2013,Nam2016}. Furthermore, giant Rashba SOC and 2D superconductivity were found in the heterostructure Tl$_3$Pb/Si(111) \cite{Matetskiy2015}, which possesses the same lattice structure of Pb$_3$Bi/Ge(111), as shown in Fig.~\ref{fig1:structure}(a). The microscopic mechanism of superconductivity in these systems may depend on their band fillings, leading to different DOS and electron correlations. To simplify the description, in the present study, we capture the superconductivity observed in these experiments by employing the local {\it s}-wave pairing as
\begin{equation}
H_s=\Delta \sum_{i\sigma}a_{i\sigma}^\dagger a_{i\bar{\sigma}}^\dagger + \Delta  \sum_{l\sigma} b_{l\sigma}^\dagger  b_{l\bar{\sigma}}^\dagger + \text{h.c.},
\label{eq:Hs}
\end{equation}
where $\sigma$ denotes spin up($\uparrow$) or down($\downarrow$), $\bar{\sigma}$ is reversed $\sigma$, and $\Delta$ is the superconducting gap. Hereafter, we choose $\Delta=0.3$ meV to be comparable with experiments\cite{Yamada2013,Zhang2010}. In order to drive this system into the topological region, a perpendicular Zeeman field is further introduced by
\begin{equation}
H_z=V_z \sum_{i} a_{i}^\dagger \sigma_z a_{i}+V_z\sum_{l} b_{l}^\dagger \sigma_z b_{l},
\label{eq:Hz}
\end{equation}
where $V_z$ represents the Zeeman energy. In Eqs.~(\ref{eq:Hs}) and (\ref{eq:Hz}), for simplicity, we choose the same superconducting gap $\Delta$ and Zeeman energy $V_z$ for the Pb and Bi atoms. In the presence of an external magnetic field $\bm{B}$, the Zeeman energy is $V_z=-g\mu_B \bm{S} \cdot \bm{B}/\hbar$, where $g$ and $\bm{S}$ are respectively the Land\'{e} {\it g}-factor and spin angular momentum. Since heavier metals usually possess larger {\it g}-factors, for example $g\sim100$ for Pb thin films \cite{Lei2018}, the Zeeman energy of our system can be estimated as $V_z = 10\Delta = 3$ meV by taking account of $g = 100$ and an externally applied magnetic field of 1T. In the momentum space, the Bogoliubov–de Gennes (BdG) Hamiltonian of $\mathcal{H}(\bm{k})$ can be expressed as 
\begin{equation}
\mathcal{H}_{\text{BdG}}(\bm{k})= 
\begin{bmatrix}
\mathcal{H}(\bm{k}) & \Delta(\bm{k}) \\
\Delta^{\dagger}(\bm{k}) & -\mathcal{H}^{\text{T}}(-\bm{k})
\end{bmatrix},
\label{HBdG}
\end{equation} 
with
\begin{equation}
\Delta(\bm{k}) = 
\begin{pmatrix}
0 & \Delta \\
-\Delta  & 0 \\
\end{pmatrix},
\end{equation}
where the Zeeman energy term has been incorporated into the normal-state Hamiltonian $\mathcal{H}(\bm{k})$. 

Based on the symmetry classification of topological materials \cite{Schnyder2008,Kitaev2009,Ryu2010a,Chiu2016}, the present system belongs to class-$D$ due to the absence of time-reversal symmetry, thus its corresponding topological invariant is the first Chern number. Given the very complex band structures in Fig.~\ref{fig1:structure}(c), it is extremely time-consuming to get a convergent Chern number via Kubo formula calculations \cite{Thouless1982}. Here we employ an alternative approach, which largely simplifies the numerical calculation of the Chern number for a complicated system\cite{Fukui2005}. The first step is to discrete the BZ into an uniform mesh, marking each unit plaquette using $\bm{k}_{i,j}=(k_i, k_j)$, with $ 0 \leq i,j \leq N-1$ and the relationship between i and j is depend on the shape of the BZ. Different from the Berry connection, we then define a new variable using the gauge-dependent wave function of the {\it n}-th band $\ket{n(\bm{k})}$ at each plaquette as
\begin{equation}
U_{\mu}(\bm{k}_{i,j}) = \frac{\braket{n(\bm{k}_{i,j})|n(\bm{k}_{i+\delta_{\mu,1},j+\delta_{\mu,2} })}} {\lvert\braket{n(\bm{k}_{i,j})|n(\bm{k}_{i+\delta_{\mu,1},j+\delta_{\mu,2} })}\rvert},
\end{equation}
in which $\mu = 1,2$. Starting from this variable, a gauge-invariant lattice field can be further written by
\begin{equation}
F_{ij} = \ln U_{1}(\bm{k}_{i,j}) U_{2}(\bm{k}_{i+1,j}) U_{1}(\bm{k}_{i,j+1})^{-1} U_{2}(\bm{k}_{i,j})^{-1},
\end{equation}
and is connected to the Chern number of the {\it n}-th band by summarizing the lattice field across the whole mesh through
\begin{equation}
c_{n} = \frac{1}{2 \pi i} \sum_{i,j} F_{ij}.
\end{equation}
Using the scheme described above, we are capable of acquiring a convergent Chern number with a relatively small $N$. Combining the Chern number with the quasiparticle band gap derived from the BdG Hamiltonian $\mathcal{H}(\bm{k})$, we are ready to investigate the topological phase transition in the Pb$_3$Bi/Ge(111) system.

\begin{figure}[tb]
	\includegraphics[width=1\columnwidth]{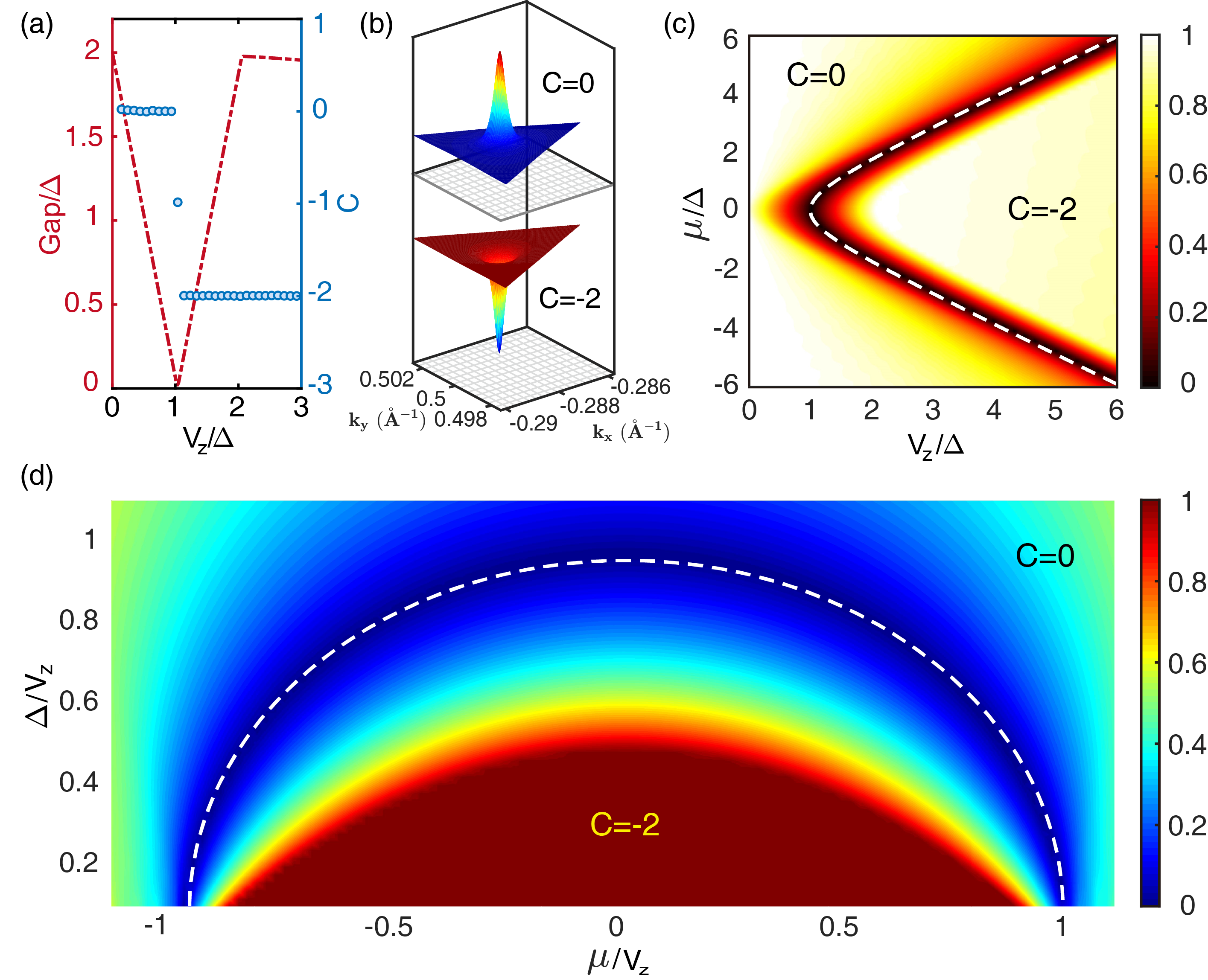}
	\caption{(a) Quasiparticle band gap (red) and Chern number (blue) as functions of $V_z$ with $\Delta =0.3$ meV. The chemical potential $\mu = 0$ meV is set to the Dirac point at K. (b) Berry curvature around the K point before (upper) and after (lower) the topological phase transition. (c) Quasiparticle band gap as a function of $\mu$ and $V_z$, where $\Delta =0.3$ meV, and the gap is normalized by $2\Delta$. (d) Quasiparticle band gap as a function of $\mu$ and $\Delta$ with $V_z =0.5$ meV, where the gap is also normalized as in (c). The phase boundary in both (c) and (d) are highlighted by dashed white curves.}
	\label{fig2:chern}
\end{figure}

As shown in Fig.~\ref{fig2:chern}(a), the quasiparticle band gap decreases linearly with increasing $V_z$, closes at $V_z\sim\Delta$, and reopens when $V_z>\Delta$. The corresponding Chern number changes from $\mathcal{C}=0$ to $\mathcal{C}=-2$, suggesting a topological phase transition at the gap closing point. This topological phase transition is different from that proposed to happen around the $\Gamma$ point in the earlier studies, where $\mathcal{C} =\pm1$ \cite{Sau2010,Lutchyn2010,Alicea2010}. The reason for such a high Chern number topological superconducting state in our system is rooted in the particular lattice symmetry. The gapless Dirac cone at the K point can be viewed as a magnetic monopole, and a band inversion will reverse its sign, leading to an abrupt change of the Chern number $\delta \mathcal{C} = \pm1$. For the present system, the Berry curvature around K and K' are the same due to the mirror symmetry. Therefore, upon a band inversion, the Berry curvature around these two points will reverse their sign, as demonstrated in Fig.~\ref{fig2:chern}(b). Meanwhile, the total Chern number $\mathcal{C}$ will transform into $\pm 2$, with the sign depends on the direction of the applied magnetic field.

In Figs.~\ref{fig2:chern}(c) and (d), we plot the quasiparticle band gap upon varying the chemical potential $\mu$, Zeeman energy $V_z$, and superconducting gap $\Delta$. The topological phase space of $\mathcal{C}=-2$ is quite large when the Fermi level is tuned to be around the Dirac point, providing a broad parameter window for realizing TSC with a minimally required Zeeman energy $V_z \sim \Delta $. Moreover, the topological phase boundary is well described by $V_z = \sqrt{\Delta^2+\mu^2}$, a criterion derived around the $\Gamma$ point \cite{Sau2010,Sau2010a}. This is because the topological properties of the low-energy Dirac cones around the K and $\Gamma$ points are very similar. In addition, the present system is also an appealing platform for realizing TSC around the $\Gamma$ point, since the Fermi level is at the corresponding Dirac point even without doping.
 
\begin{figure*}[tb]
	\includegraphics[width=1.3\columnwidth]{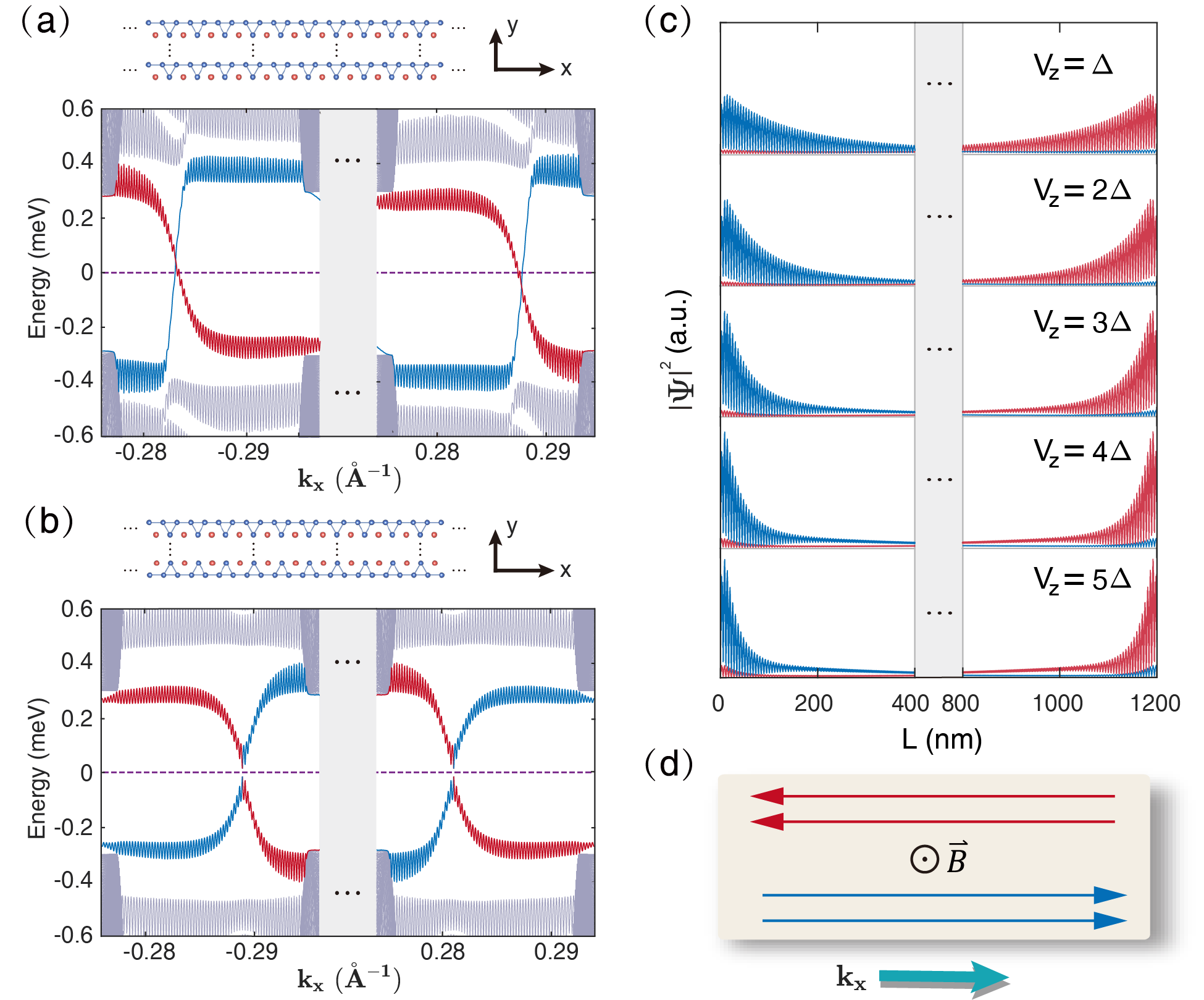}
	\caption{(a)-(b) Energy spectra for the corresponding ribbon structures, as shown in the upper panels. For numerical calculations, we choose 2000 unit cells, equaling to 1200 nm long in the $y$ direction and apply periodic boundary condition in the $x$ direction. To be specific, we use a superconducting gap $\Delta = 0.3$ meV and a Zeeman energy $V_z=5\Delta=1.5$ meV. The red and blue colored chiral Majorana edge modes originate from the two edges, propagating along opposite directions. (c) Real-space distributions of the zero-energy chiral Majorana edge modes corresponding to the energy crossing point at k$_x$=0.288{\AA} in (b), where their respective Zeeman energies chosen are given at the upper right corners. (d) Schematic diagram of the zero-energy chiral Majorana edge modes in Pb$_3$Bi/Ge(111).}
	\label{fig3:edge}
\end{figure*}

\subsection{Chiral Majorana edge modes}
\label{sec:vhs}

Bulk-boundary correspondence plays a crucial role in examining the topologically nontrivial properties of realistic materials \cite{Mong2011}. For the class-$D$ TSCs, this correspondence predicts that the number of chiral edge modes is equal to the first Chern number of its bulk state. To connect our theoretical predictions to potential experimental realization, we investigate the chiral Majorana edge modes in Pb$_3$Bi/Ge(111) hereafter.

Employing a cylinder geometry with periodic boundary condition in the x direction and open boundary condition in the y direction, we construct the Hamiltonian of the Pb$_3$Bi/Ge(111) nanoribbon and explore the edge state properties under different boundary configurations. Two typical ribbon structures shown in the upper panels of Figs.~\ref{fig3:edge}(a) and (b) are considered, where the latter one possesses in-plane inversion symmetry. Their corresponding quasiparticle energy spectra are depicted in the lower panels of Figs.~\ref{fig3:edge}(a) and (b), where both of them exhibit two chiral Majorana edge modes propagating along the same direction at one edge, consistent with $\mathcal{C} = -2$. The difference between Figs.~\ref{fig3:edge}(a) and (b) is that the chiral Majorana edge modes from the two edges intersect at a finite energy in (a), while at zero energy in (b). This phenomenon can be understood by the following symmetry arguments. On the one hand, the inherent particle-hole symmetry of the BdG Hamiltonian is expressed as  $P\mathcal{H}_{\text{BdG}}(k_x)P^{-1}=-\mathcal{H}_{\text{BdG}}(-k_x)$, indicating $E(k_x) = -E(-k_x)$. This feature is evident in both of the quasiparticle energy spectra, as shown in Figs.~\ref{fig3:edge}(a) and (b). On the other hand, the in-plane inversion symmetry implies that $E(k_x) = E(-k_x)$. Therefore, the zero-energy intersections of the chiral Majorana edge modes in Fig.~\ref{fig3:edge}(b) is protected by the combination of particle-hole symmetry and in-plane inversion symmetry.

In our present calculations, the chiral Majorana edge modes possess noticeable oscillations that can be attributed to strong finite size effects \cite{Zhou2008,Wada2011}. In Fig.~\ref{fig3:edge}(c), we plot the real-space distributions of these edge states at the zero-energy intersection point k$_x$=0.288{\AA} in Fig.~\ref{fig3:edge}(b). As expected, the states from the two edges are symmetric due to in-plane inversion symmetry and decrease exponentially into the bulk. Moreover, the penetration length $\xi$ of these chiral edge states defined by real-space distribution $ \phi = e^{-x/\xi}$ decreases with increasing $V_z$, pointing to the more localized zero-energy edge modes for larger $V_z$. In comparison to the earlier reports based on proximity effects\cite{Qi2010}, edge modes in our system are more dispersive in real space due to the smaller superconducting gap. Based on the above results, a schematic diagram that emphasizes the two zero-energy chiral Majorana edge modes in Pb$_3$Bi/Ge(111) is depicted in Fig.~\ref{fig3:edge}(d).

\section{Discussion}
\label{sec:discussion}

In this section, we discuss the potential experimental realization of the predicted TSC and chiral Majorana edge modes in Pb$_3$Bi/Ge(111). The gapless Dirac cone at the K point is located $\sim 80$ meV below the Fermi level as shown in Figs.~\ref{fig1:structure}(b) and (c), and the calculated carrier density required to move the Fermi level to this point is $\sim1.0 \times 10^{14}$ cm$^{-2}$, which is accessible, for example via ionic liquid gating\cite{Ye2012}. Moreover, since the experimentally reported superconducting gaps for Pb thin films and related systems range from 0.3 meV to more than 1 meV,\cite{Qin2009,Zhang2010,Sekihara2013,Yamada2013,Nam2016} we choose the lower bound of $\Delta = 0.3$ meV as an illustration in the present study. Meanwhile, the observed critical perpendicular magnetic fields $H_{c\bot}$ in these systems range from 0.15 T to over 1 T,\cite{Zhang2010,Yamada2013} indicating that superconductivity can survive in the presence of a sufficiently strong field. As we have noted in the past section, the Land\'e {\it g}-factor is expected to be large enough and enable the system to become a TSC under a relatively moderate magnetic field that weakens but does not completely destroy the superconductivity. Specifically, by choosing a proper value of $g\sim100$,\cite{Lei2018} the topological phase transition takes place at $V_z \sim \Delta$, which can be reached by applying a magnetic field of $0.1 $ T.

On the strength of the above analysis, we are capable to conceive a practical experimental architecture to realize our predictions. There are two typical approaches for introducing a Zeeman field into a superconducting system. The first is via magnetic proximity effects, realized by placing magnetic materials close to a superconductor\cite{Yu2005,Shiba1968,Rusinov1969}. By employing this scheme, earlier experimental studies have reported signatures of Majorana zero modes in related systems, such as the magnetic islands in proximity to a superconducting Pb thin film\cite{Menard2017,Menard2019} or Re surface\cite{Palacio-Morales2019}. One disadvantage of this approach is that the Zeeman field is difficult to modify, because the magnetic material is usually deposited on the substrate. The other approach is directly applying a perpendicular magnetic field\cite{Xu2015,Sun2016}, which we choose here to achieve a higher tunability. After the Zeeman splitting energy $V_z$ is introduced into our system, the penetration length $\xi$ of the Majorana zero modes in a magnetic vortex can be approximated by $\xi=a\lambda/(V_z-\Delta)$ \cite{Sato2009} when $V_z$ is not too large, where $a$ is the lattice constant and $\lambda$ is the Rashba coefficient. The large {\it g}-factor of the present system ensures that $\xi$ can be manipulated substantially via an external magnetic field. As depicted in Fig.~\ref{fig3:edge}(c), for an experimentally realizable magnetic field $B=0.5$ T, the penetration length of the zero-energy Majorana edge modes can be reduced to $\sim 50$ nm. Therefore, we propose that a Pb$_3$Bi superconducting island with a size of a few hundred nanometers can harbor zero-energy Majorana edge modes and may be tested by low-temperature scanning tunneling microscopes\cite{Nadj-Perge2014,Feldman2017}. Similarly, the condition for observing the Majorana zero modes in a magnetic vortex is that the vortex separation is larger than $2\xi \sim 100$ nm under the applied magnetic field of $B=0.5$ T.

\section{Summary}
\label{sec:dis}

In summary, we have developed a phenomenological tight-binding model for the heterostructure Pb$_3$Bi/Ge(111) and shown its ability to capture the main features of the giant Rashba splitting and type-\uppercase\expandafter{\romannumeral2} VHS obtained from DFT calculations. Besides the time-reversal invariant $\Gamma$ and M points, we have also identified a gapless Dirac cone protected by the $C_{3v}$ lattice symmetry at the K point. By tuning the Fermi level to this symmetry-protected Dirac point and considering the {\it s}-wave superconductivity of the parent system, we have illustrated that this system can be transformed into a topological superconducting phase with the first Chern number $\mathcal{C} = -2$ in the presence of a sufficiently strong Zeeman field. Through the edge state calculations, we have further demonstrated two chiral Majorana edge modes propagating along the same direction of the system with proper boundaries, corresponding to the nontrivial topology of $\mathcal{C} = -2$. By evaluating and specifying the physically realistic conditions, including the superconducting gap, Land\'{e} {\it g}-factor, and critical magnetic field, we have put forward a practical experimental architecture for future validation of the predictions made here. Our central findings are useful for realizing chiral topological superconductivity and observing Majorana edge modes in 2D or related interfacial systems.

\vspace{-0.2cm}
\section*{ACKNOWLEDGMENTS}
\vspace{-0.2cm}
\label{sec:ack}
This work was supported by the National Key R\&D Program of China (Grant No. 2017YFA0303500), the National Natural Science Foundation of China (Grant Nos. 11634011, 11722435, 11974323, and 11904350), the Anhui Initiative in Quantum Information Technologies (Grant No. AHY170000), and the Strategic Priority Research Program of Chinese Academy of Sciences (Grant No. XDB30000000).

\bibliography{main} 



\end{document}